# Absolute Determination of Optical Constants by a Direct Physical Modeling of Reflection Electron Energy Loss Spectra

H. Xu,[1] B. Da,[1] J. Tóth[3], K. Tőkési[3,4] and Z.J. Ding[1,2 a)]

[1]*Hefei National Laboratory for Physical Sciences at Microscale and Department of Physics,*

*University of Science and Technology of China, Hefei 230026, Anhui, P.R. China*

[2]*Key Laboratory of Strongly-Coupled Quantum Matter Physics, Chinese Academy of Sciences*

[3]*Institute for Nuclear Research, Hungarian Academy of Sciences (ATOMKI), P.O. Box 51,*

*Debrecen, Hungary*

[4]*ELI-ALPS, ELI-HU Non-profit Ltd., Dugonics tér 13, H-6720 Szeged, Hungary*

We present an absolute extraction method of optical constants of metal from the measured reflection electron energy loss (REELS) spectra by using the recently developed reverse Monte Carlo (RMC) technique. The method is based on a direct physical modeling of electron elastic and electron inelastic scattering near the surface region where the surface excitation becomes important to fully describe the spectrum loss feature intensity in relative to the elastic peak intensity. An optimization procedure of oscillator parameters appeared in the energy loss function (ELF) for describing electron inelastic scattering due to the bulk- and surface-excitations was performed with the simulated annealing method by a successive comparison between the measured and Monte Carlo simulated REELS spectra. The ELF and corresponding optical constants of Fe were obtained from the REELS spectra measured at incident energies of 1000, 2000 and 3000 eV. The validity of the present optical data has been verified with the *f*- and ps-sum rules showing the accuracy and applicability of the present approach. Our data are also compared with previous optical data from other sources.

There is a continuous interest and effort on the determination of optical constants of solids due to their importance in both fundamental researches and applications. Optical methods based on reflectance and absorption spectroscopy with ellipsometry were extensively employed up to now and the measured data for metals and semiconductors were compiled to form a database of optical constants.[1,2] However, many materials still lack the data in the intermediate photon energy range around 20~50 eV. Furthermore, the available data usually consist of different energy regions measured by different groups and means and,

---

a) e-mail: zjding@ustc.edu.cn

therefore, the data may not be smoothly joined. On the other hand, the electron energy loss spectroscopy[3-5] can provide alternative way for deriving information of dielectric response of solid to external electric field carried by electrons, which is, in principal rather different technique compared with optical methods. In recent years a technique based on the reflection electron energy loss spectroscopy (REELS) has been developed[6-16] to obtain optical constants in a rather wide range of energy loss of electrons (i.e. photon energy). The typical energy loss range is between 1 and 100 eV, while the measurements can be performed in once or may be several times under different experimental conditions but with the same spectrometer. Such an ability to derive optical constants in a wide photon energy range with only one spectrum is the main advantage of REELS comparing with the optical measurements. In addition, it also holds the opportunity to get the optical constants for nonzero momentum transfers.

In deriving the energy loss function (ELF), $\text{Im}[-1/\varepsilon(\omega)]$, and thereby the optical constants $(n,k)$, where $\varepsilon = n + ik$ is complex dielectric function of the solid, from the measured REELS spectra the precise and accurate knowledge of the electron energy loss processes, i.e. the combination of multiple elastic scattering with the bulk- as well as surface-inelastic scattering of electrons interacting with the sample has a crucial importance. Aiming at extracting from REELS spectrum Many of the previous works[6-8,16] used an analytical algorithm[17] to get the single inelastic scattering distribution, i.e. the differential inverse inelastic mean free path (DIIMFP), by neglecting completely the influence of the elastic scattering. The calculation procedure was later modified to include the effect of elastic scatterings by applying a scaling factor[9,10], however, the obtained effective ELF has ambiguous physical meaning. In the same spirit, a REELS spectrum was analytically described as a convolution of multiple inelastic scattering contributed from surface and bulk excitations.[11,12,18] The weighting factors for the corresponding energy loss distributions represent only the partial intensity of electrons inelastically scattered in the solid. Although these attempts have promoted advances in understanding of electron interactions with solid surfaces and provided valuable optical data for some metals,[12] such analytical modeling has still serious problems: a) Calculations require pre-knowledge as input of inelastic mean free path (IMFP) and surface excitation parameter (SEP); b) Despite that the REELS spectrum intensity is scaled with the elastic peak intensity, without taking account of elastic scattering in the determination of optical constants the obtained data is not absolute, and therefore the ELF must be scaled by introducing artificial scaling factors;[12] c) Furthermore, the shape of the REELS spectrum is actually also sensitive to the ratio of cross sections between elastic scattering and inelastic scattering, and thus it has influence to the derived ELFs. d) Last but not least, the analytical algorithm[12] which includes surface excitation assumes homogenous scattering properties of a sample while the surface excitation is in fact depth dependent.[19-21]

A significant improvement has been recently achieved based on a new numerical modelling,[15] namely the so called reverse Monte Carlo (RMC) method which integrates Monte Carlo (MC) simulation of REELS spectrum and Marokv chain Monte Carlo (MCMC) technique for updating of ELF during the calculation. Our MC simulation of electron trajectories is fully based on our knowledge of physical process of electron-sample interaction and is the one of the most powerful numerical technique to surface analysis.[22] Employing the up-to-date MC simulation of electron scattering in the surface region the RMC method overcomes the drawbacks of an analytic method: a) Elastic scattering of electrons is taken into account completely, which ensures the absolute determination of the ELF values rather than a relative one. B) The multiple scattering effects including the surface excitation are taken into account in a well-developed MC technique[21] which has been proven to be the most accurate way so far for REELS spectrum analysis. We note, however, that such a physical model based MC simulation is quite time consuming and impractical to extract the optical data from comparison between measured spectrum and simulation. C) Therefore a probabilistic searching technique, the simulated annealing (SA) method[23] as one of MCMC methods, is employed to build a fast adapting procedure of ELF values through the optimization of oscillator parameters. The principle of SA method here is to find the global minimum of a potential function, defined as the overall difference between the measured and simulated REELS spectra, in oscillator parameter space.

Although absolute values of optical constants were obtained successfully by the RMC method for $SiO_2$,[15] it still needs further improvements for elemental solids with complex electronic structures, such as transition metals. This is because the previous simplification, i.e. depth independent surface excitation, made in the calculation of DIIMFP becomes poor and limits quantitative evaluation of the characteristics of electronic excitation. In the present work, the framework of the RMC method is extended to allow the MC modelling of depth dependent surface excitations by integrating a more sophisticated approach to electron inelastic scattering. The surface mode of collective excitations[24] is due to the presence of sample boundary between material and vacuum. The real measurable electron energy loss spectrum is a superposition of energy loss spectra excited either in the bulk or in the surface region and therefor it must be decomposed. Although SEP, defined as mean number of surface plasmons excited by electrons moving across a solid surface, which can be obtained by integrating the surface component in DIIMFP, may be useful to investigate the effects of surface excitation in some studies;[25-27] but for an accurate MC simulation of REELS spectra we must use directly a spatial (depth and directional) dependent DIIMFP. Two typical models are usually apply to calculate the DIIMFP, i.e. the semi-classical one[20,28] and quantum mechanical one[29,30], respectively. In our recent work the semi-classical model is used, partly because it gives, in most cases[31], very close results with the quantum mechanical model, and partly and more importantly, because it is computationally more efficient than the quantum mechanical one.

In our present procedure of RMC, a trial ELF is parameterized as sum of a number of Drude-Lindhard functions:

$$\mathrm{Im}\left[\frac{-1}{\varepsilon(q,\omega)}\right] = \sum_{i=1}^{N} A_i \, \mathrm{Im}\left[\frac{-1}{\varepsilon(q,\omega;\omega_{pi},\gamma_i)}\right], \quad (1)$$

where the 3$N$ oscillator parameters, $A_i$, $\omega_{pi}$ and $\gamma_i$ are, respectively, the oscillator strength, energy and width of the $i$-th oscillator. They are initially arbitrarily selected at long wavelength limit $q \to 0$. Based on the first set of the oscillator parameters, a MC simulation is performed to produce an initial simulated REELS spectrum, $I_0^{sim}(\Delta E)$. A "goodness" of the oscillator parameters after $n$-iteration in SA is defined as,

$$\chi_n^2 = \sum_j \left(I_n^{sim}(\Delta E_j) - I^{exp}(\Delta E_j)\right)^2 \Big/ T(\Delta E_j) \quad (2)$$

where $I^{exp}(\Delta E_j)$ is the experimental intensity at each energy loss values of $\Delta E_j$. Numerator and denominator in Eq. (2) are respectively the "potential energy" and the "temperature" in SA appearing in Boltzmann distribution. Here $T$ is the actually a weighting factor to specify the importance of the specific energy loss zone, and it can be assigned different values for different zones. The optimal values of 3$N$ oscillator parameters are calculated using the successive approximation, ie. the simulation of REELS spectrum calculation is performed iteratively in order to minimize goodness parameter, $\chi_n^2$. In this case the obtained ELF will converge and represent the true value of the sample[15]. For materials having a complex shape of ELF, like transition metals, about fifty or more Drude-Lindhard terms are necessary for an adequate expression of ELF. The determination of the ELF thus actually turns into a task of global optimization in a hyperspace of over one hundred dimensions. MCMC sampling, i.e. Metropolis importance sampling, and SA algorithm for adjusting the parameter set is necessary to reduce computation time. The optimized final ELF as a sum of Drude-Lindhard terms is independent of the choice of initial parameter set, as a consequence of the basic property of the MCMC method.

Three REELS spectra for a mechanically polished polycrystalline iron sheet sample were recorded with primary electron energies at 1000, 2000 and 3000 eV in an energy loss range of 0-100 eV by a home built electron spectrometer of ESA-31 in ATOMKI.[32] The analyzer works in a fixed retardation ratio mode with a relative energy resolution of $5\times10^{-3}$. In the present experiments the used pass energies were around 100 eV, and in this way the analyzer energy resolution was around 0.5 eV. The full widths at half maximum, being the convolution of the analyzer and the electron source generated widening of the elastic peak, were around 0.6-0.7 eV. The incident angle of the primary electron beam is 50° with respect to the surface normal of the sample and the angle of the analyzed electrons is 0° relative to the surface normal. MC simulations for the REELS spectra were performed for this geometry and for the three primary electron energies.

Fig. 1 shows the evolution of ELF as a function of the number of MCMC steps. The corresponding goodness parameter, or equivalently the potential surface in the parameter space is changed for each trial ELF with each steps. Although it is fluctuated step by step it always approaches to the steady minimum. For an intuitionistic view, evolution of the desired optical constants and the IMFPs are also calculated and plotted corresponding to the updating ELFs.

Each REELS spectrum at certain experimental condition yields one ELF of the sample. We have performed the RMC calculations to get ELFs for three energies to check their consistency. Fig. 3(a) shows that the agreement on the absolute intensity scale between the final simulated and experimental REELS spectra from elastic peak down to energy loss of 100 eV at primary electron energies of 1000, 2000 and 3000 eV are all excellent. Hhere the intensity is scaled with that of elastic peak as shown by the inset of Fig. 3(a). To reveal the importance of surface excitations, we have calculated REELS spectra contributed only by pure bulk excitations. The contributions of surface excitations in this case are decomposed by subtraction the pure bulk contributions from the full spectra, as displayed in Fig. 3(a). The corresponding ELFs, for the three energies obtained directly by RMC in absolute values and without resorting to any normalization procedures are illustrated in Fig. 3(b), together with corresponding surface ELFs. We found an overall consistency as expected. To check the accuracy of these ELFs, $f$- and ps-sum rules were calculated, which are defined respectively by,

$$Z_{eff} = \frac{2}{\pi \Omega_p^2} \int_0^\infty \omega \, \text{Im}\left[-1/\varepsilon(\omega)\right] d\omega; \qquad (3)$$

$$P_{eff} = \frac{2}{\pi} \int_0^\infty \frac{1}{\omega} \text{Im}\left[-1/\varepsilon(\omega)\right] d\omega, \qquad (4)$$

where $\hbar\Omega_p = \sqrt{4\pi n_a e^2 / m_e}$ and $n_a$ is the atomic density sample. The obtained results are summarized in Table I, where the data above 100 eV are taken from measurements performed by Henke.[33] The nominal theoretical values for $f$- and ps-sum rules are the atomic number (Z=26 for Fe) and 1, respectively. We found that for all three energies the obtained values are very close to the nominal theoretical values and the relative errors are very small for both sum rules. This clearly indicates that our calculated ELFs gives reasonable optical properties of Fe sample from the near visible to the soft x-ray photon energy region. Therefore this also confirmed that our MC simulation with a combination of inelastic scattering cross section calculated by a semi-classical dielectric functional approach and Mott's cross section for elastic scattering describe accurately the electron transport processes. The sum rules calculated taking average over the three energies, are also given in Table I.

To minimize the uncertainties, we take an average over the ELFs for the three energies. Fig. 4 shows the averaged ELF in comparison with the results of Werner,[12] Palik[1] and Henke.[33] In the low energy loss region (less than 25 eV), the RMC result agrees well with the Werner's data from a deconvolution of REELS spectra while it deviates from Palik's data above 15 eV. In the intermediate energy region, between 25 and 50 eV, our RMC result is more close to Werner's DFT calculation and it deviates from Henke's data deduced from atomic scattering factors below 40 eV. In the high energy region, the present ELF closes to the Henke's data but has sharper $M_{2,3}$-edge around 55-60 eV and it is not as strong as that of Werner's data by REELS measurement and DFT calculation. We note that the agreement with Henke's data in the high energy region is important because the absorption properties of a solid should approach to the atomic properties above the ionization edge. Thus the present absolute ELF generally falls into the data distribution range from different sources and it agrees with REELS data in low energy region, with the DFT calculation in intermediate energy region and with the atomic data in high energy region.

Finally, optical constants are obtained from ELF according to the Kramers-Kronig relation.[5] Fig. 5 shows the calculated refractive index *n* and extinction coefficient *k* of Fe in the photon energy range between 0-100 eV in comparison with the widely used Palik's database of optical constants. Furthermore, IMFP, one of the most important parameters for chemical quantification by surface electron spectroscopy techniques, is calculated by a dielectric response theory[34] and compared with the NIST data[35] in Fig. 6. A good agreement with TPP-2M formula[36] and the elastic peak electron spectroscopy (EPES) measurements by Lesiak et al.[35] demonstrates the ability of the present RMC method for the determination of IMFP from REELS spectra. Note that, by virtue of a comprehensive description of experiment, REELS excels EPES in that the IMFP can be fundamentally deduced in the whole energy range with one measurement of REELS spectra at one primary energy rather than separate measurements of EPES at required primary energies.

In summary, we have obtained ELF and optical constants of iron in energy range between 0-100 eV from the measured REELS spectra with the help of the recently developed RMC technique. The *f*- and ps-sum rules for the energy averaged ELF are, respectively, are 25.99 and 1.04 with relative errors of -0.038% and 4.0%. The ELF used in the REELS spectrum simulation is approximated as the sum of the Drude-Lindhard type functions whose parameters are determined by a global optimization with a MCMC method. The optimization procedure modulates the simulated REELS spectrum to approach the measured one. A combination of a spatially varying DIIMFP under semi-classical framework of dielectric functional approach for electron inelastic scattering with Mott's cross section for electron elastic scattering was found to give accurate description of electron transport in a REELS experiment. Advantage of REELS in deriving the IMFP of Fe is demonstrated by comparing the current calculation with NIST IMFP data.

This work was supported by the National Natural Science Foundation of China (Nos. 11274288 and 11574289), the National Basic Research Program of China (No. 2012CB933702), the Hungarian Scientific Research Fund OTKA Nos. NN 103279 and K103917 and by the European Cost Actions CM1204 (XLIC) and CM1405 (MOLIM).

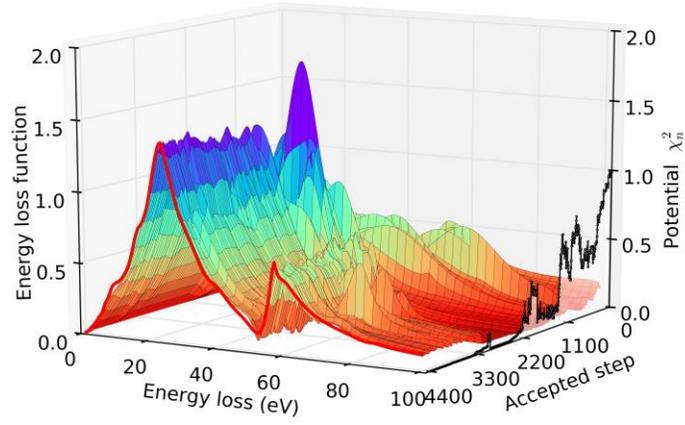

Fig. 1. Evolution of ELF in a RMC process, showing updating of the trail ELFs with MCMC iteration steps. The final ELF is depicted in red line. Variation of "potential" is displayed by black line.

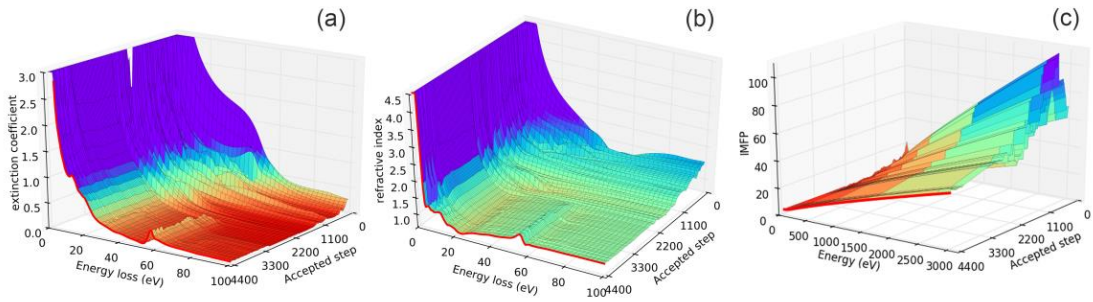

Fig. 2. Evolution of optical constants (Fig. (a). Extinction coefficient. Fig. (b) Refractive index.) and the IMFP in Fig. (c) corresponding to updating ELFs in a RMC process. The final values of each evolution is depicted in red line.

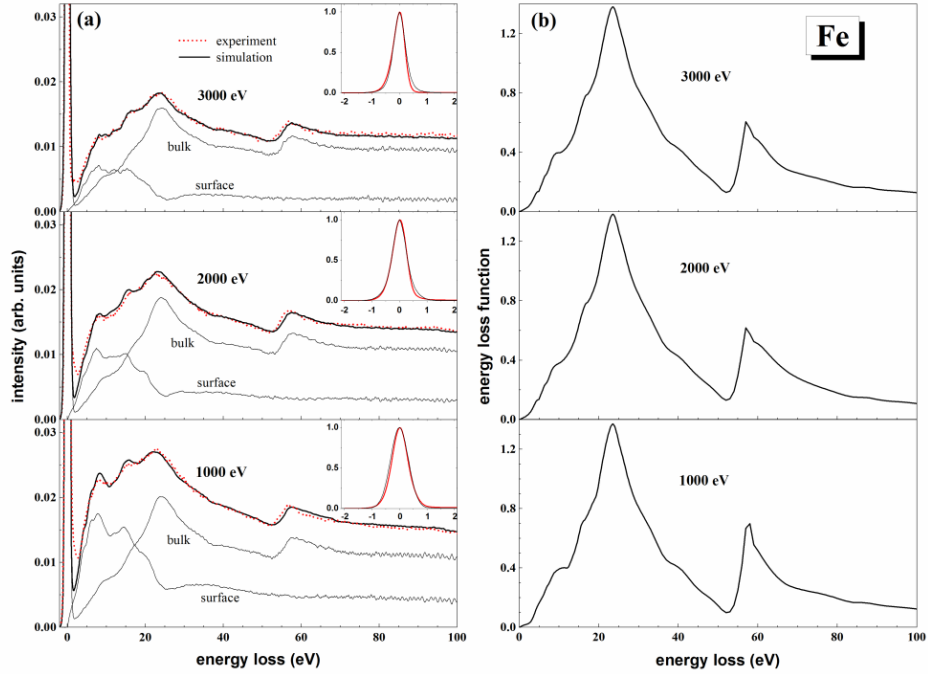

Fig. 3. (a) Comparison of simulated REELS spectra (solid lines) and measured spectra (dotted lined) of Fe at 1000, 2000 and 3000 eV. Inset shows the normalization on the elastic peak intensity. Contributions from bulk and surface excitations from MC simulation are illustrated as well. (b) ELFs, $\text{Im}[-1/\varepsilon(\omega)]$, obtained from the REELS spectra for the corresponding energies by the RMC method.

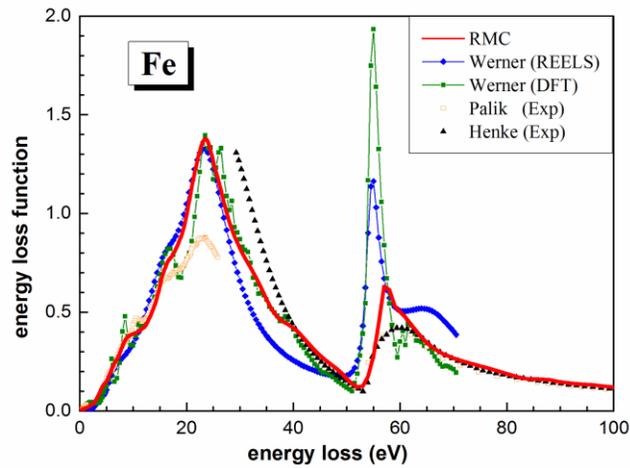

Fig. 4. Comparison of ELFs deduced by the present RMC method with Werner's REELS data and DFT calculation, Palik's compiled data (lacking data in 26-50 eV) and Henke's experiment.

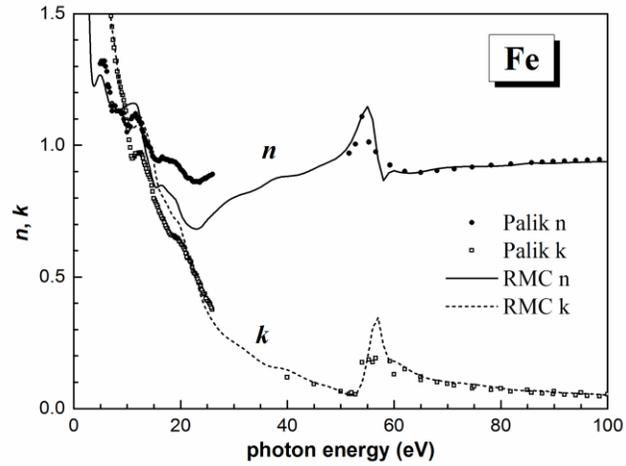

Fig. 5. Optical constants, the refractive index *n* and extinction coefficient *k*, of Fe obtained by RMC method and comparison with Palik's data.

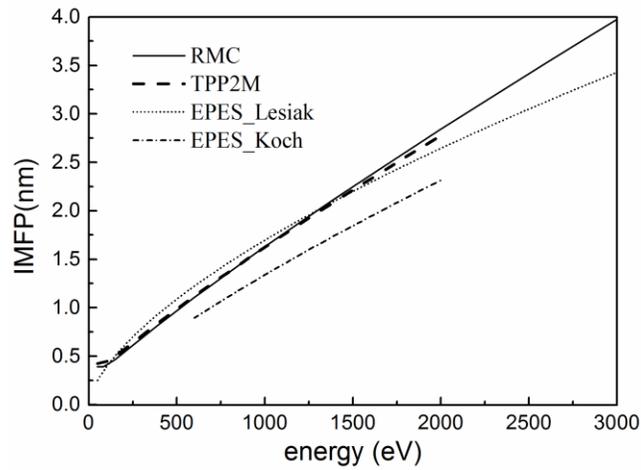

Fig. 6. Comparison on IMFP as a function of energy between present REELS analysis by RMC method (solid line), TPP-2M formula (dashed line) which coincide with the data of Kwei[34] and Gries[37] (not shown here), EPES measurements[35] by Lesiak (dotted line) and by Koch (dot and dash line).

TABLE I. List of *f*-sum and ps-sum rule checks for ELFs derived from REELS spectra at 1000, 2000 and 3000 eV and for averaged ELF (RMC)

| primary energy | $Z_{eff}$ (*f*-sum) | relative error | $P_{eff}$ (ps-sum) | relative error |
|---|---|---|---|---|
| 1000 eV | 25.95 | -0.19% | 1.024 | 2.4% |
| 2000 eV | 25.80 | -0.76% | 1.045 | 4.5% |
| 3000 eV | 26.24 | 0.92% | 1.051 | 5.1% |
| RMC | 25.99 | -0.04% | 1.040 | 4.0% |